\documentclass[journal]{IEEEtran}

\usepackage{cite}
\usepackage{amsmath,amssymb,amsfonts}
\usepackage{tabulary,booktabs}
\usepackage{algorithmic}
\usepackage{adjustbox}
\usepackage{bm}
\usepackage{multirow}
\usepackage{tabularx}
\usepackage{amsmath}
\usepackage[makeroom]{cancel}
\usepackage{xfrac}
\usepackage{wrapfig}
\usepackage{graphicx}
\usepackage{textcomp}
\usepackage{multicol}
\usepackage[table]{xcolor}
\usepackage{subfigure}
\usepackage{url}
\usepackage{array}
\usepackage{makecell}
\usepackage{slashbox}
\usepackage{soul}
\usepackage{booktabs,caption}
\usepackage[flushleft]{threeparttable}
\usepackage{changepage}
\usepackage{cite}

\begin{document}
\title{\centering \Large{{Ternary-Input Binary-Weight CNN Accelerator Design 

for Miniature Object Classification System with Query-Driven Spatial DVS}}}

\author{Yuyang Li, Swasthik Muloor, Jack Laudati, Nickolas Dematteis, 

Yidam Park, Hana Kim, Nathan Chang, and Inhee Lee
\vspace{-.5in}

}


\maketitle
 \begin{abstract}

Miniature imaging systems are essential for space-constrained applications but are limited by memory and power constraints. While machine learning can reduce data size by extracting key features, its high energy demands often exceed the capacity of small batteries. This paper presents a CNN hardware accelerator optimized for object classification in miniature imaging systems. It processes data from a spatial Dynamic Vision Sensor (DVS), reconfigurable to a temporal DVS via pixel sharing, minimizing sensor area. 
By using ternary DVS outputs and a ternary-input, binary-weight neural network, the design reduces computation and memory needs. Fabricated in 28 nm CMOS, the accelerator cuts data size by $81\%$ and MAC operations by $27\%$. It achieves 440 ms inference time at just 1.6 mW power consumption, improving the Figure-of-Merit (FoM) by 7.3× over prior CNN accelerators for miniature systems. This work has been submitted to the IEEE for possible publication. Copyright may be transferred without notice, after which this version may no longer be accessible.


\end{abstract}

\begin{IEEEkeywords}
\small Convolutional neural network, hardware accelerator, dynamic vision sensor, miniature system, low power.
\end{IEEEkeywords} 
 \vspace{-.2in}
\section{\small Introduction}
\par\small Millimeter-scale systems show great potential in security\cite{intro_miniature_imager} and biomedical applications\cite{Intro_biology}. Their small size and light weight enable breakthroughs in sensing and logging critical physical data. 
Vision-based tasks have been applied in sensing systems\cite{intro_miniature_ML_imager}, which integrate image sensors and process the captured image to extract information. Fig. 1(a) shows a traffic monitoring scenario\cite{Choi2023}. Carried by mobile devices such as drones, the sensing node undertakes two tasks: detecting the vehicles with image recognition, and track the target object with the recognized vehicle's initial position.

\par\small  The primary challenge for miniature systems lies in performing high-computation operations with severely limited resources. Convolution operations, which are required by many neural network tasks, significantly slow down processing and consume considerable energy. In the context of miniature batteries, such as the Seiko MS920SE (9.5 mm diameter) with an energy capacity of 11 mAh and a maximum discharge current of 0.8 mA, both processing time and transient power are strictly constrained. To address this,\cite{DNFv1} introduces a tracking algorithm based on Dynamic Neural Field (DNF), which reduces power consumption to 1.7 mW, meeting the power limitations of miniature systems. This approach uses temporal Dynamic Vision Sensor (DVS) signals as input, which produce ternary values of +1, 0, and -1\cite{DVS_temporal}. However, temporal DVS is primarily effective for detecting objects in motion, and struggles with slow or stationary objects. Spatial DVS detects the edge of static objects and outputs the same format of image composed by +1, 0 and -1. To enable both object detection and recognition within our target system, we propose the use of spatial DVS for object recognition alongside temporal DVS for object tracking. To minimize the increased area required by adding another sensor, which is critical in our application, we introduce a shared-pixel architecture that can be reconfigured between spatial and temporal modes. This allows the system to perform object recognition and tracking using the same sensor hardware, as shown in Fig. 1(a).
\begin{figure}[t]
\centering
\scalebox{0.75}{\includegraphics{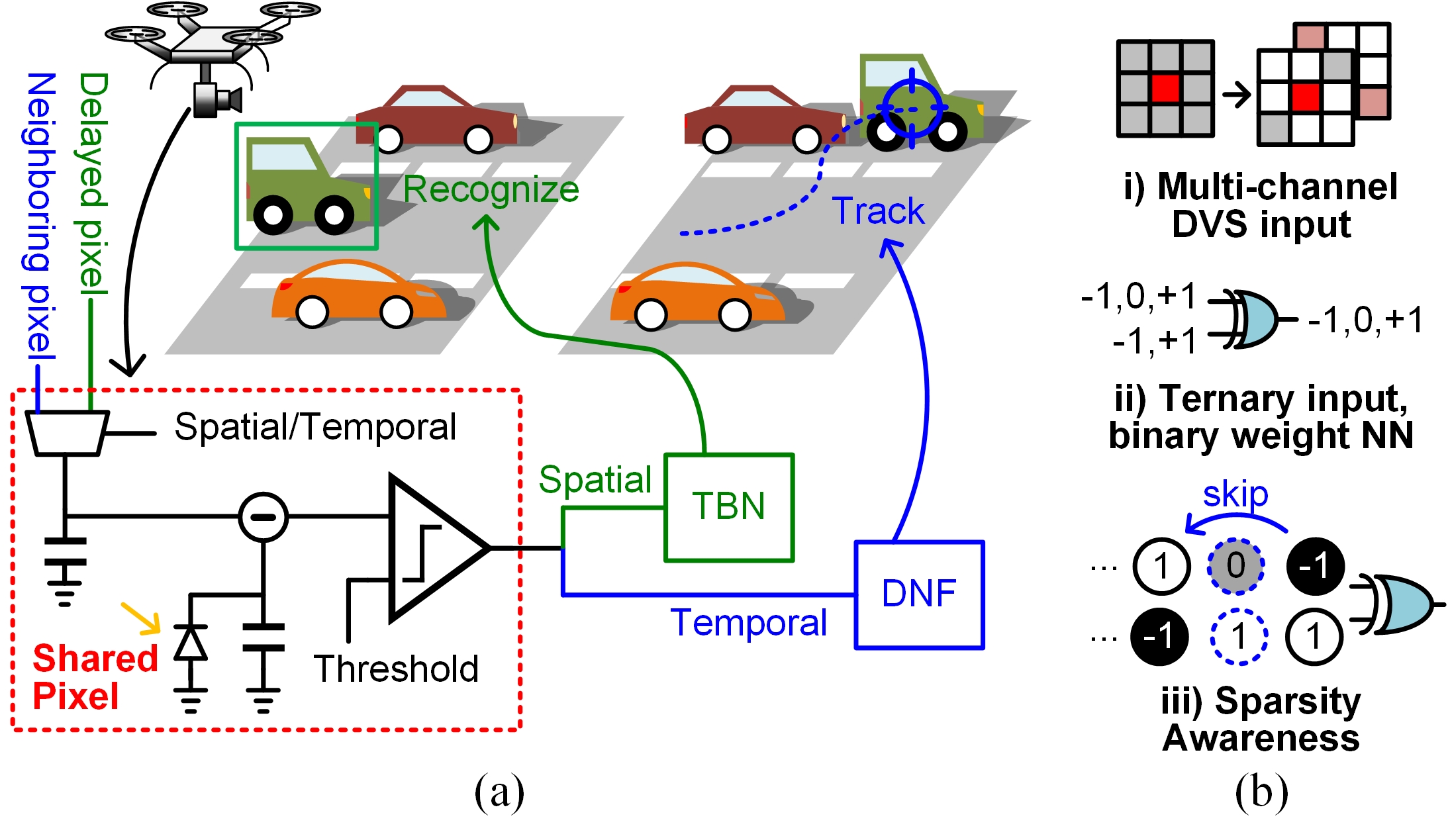}}
\DeclareGraphicsExtensions{.jpg}
\vspace{-.1IN}
\caption{Target miniature vision system. (a) Object recognition and tracking using a shared image sensor configurable as either spatial and temporal DVS. (b) Contribution of the proposed accelerator.}
\vspace{-.1IN}
\label{fig_high_level}
\end{figure}
\begin{figure*}[h]
\centering
\scalebox{0.85}{\includegraphics{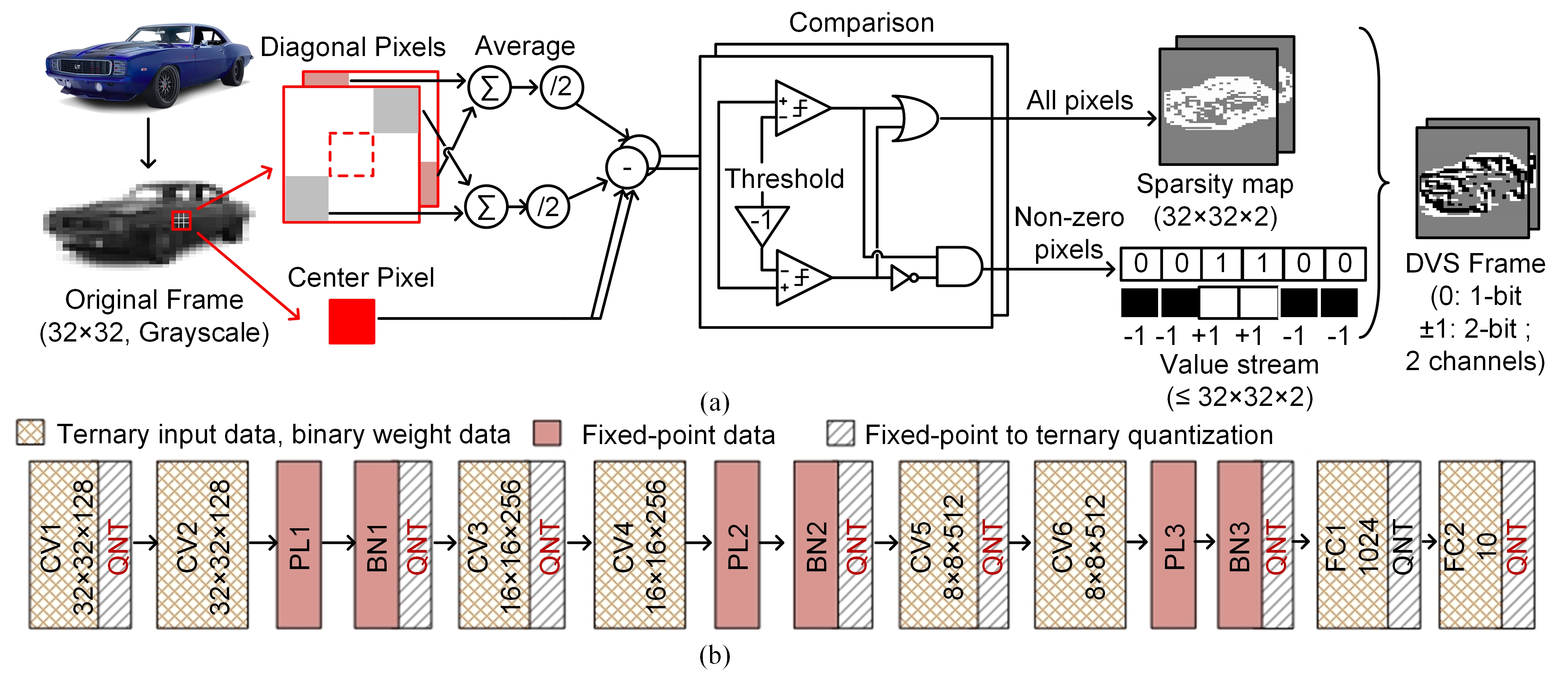}}
\vspace{-.1IN}
\DeclareGraphicsExtensions{.jpg}
\caption{\small Working mechanisms of spatial DVS. (a) Dataset generation for spatial DVS. (b) Implemented TBN architecture.}
\vspace{-.2IN}
\label{fig_DVS_TBN}
\end{figure*}

\par\small 
Neural networks (NNs) are widely used for image recognition but are often impractical for miniature systems due to the high number of multiply-accumulate (MAC) operations, leading to excessive time and power consumption. Table I summarizes the target specifications of our accelerator. To meet the constraints of a chip-layer-stacked miniature system \cite{M3}, we limit the chip size to 2.5 mm per side and the power budget to 2.4 mW, based on battery capacity. Memory is also constrained to less than 500 kB considering chip size constraint and memory capacity of prior works. For example, \cite{intro_miniature_ML_imager}integrates 426 kB of SRAM for face recognition. To stay within these limits, we use the CIFAR-10 dataset ($32\times32$ resolution). Larger datasets like ImageNet ($224\times224$) are not feasible considering the requirement of minimum. For example, A signle 128-channel convolution layer for ImageNet requires at least 1.5 MB ($244\times224\times512\times2  bits$) for input and output feature maps, resulting in 6.6 mm$^2$ in a 28 CMOS process. For processing time and recognition accuracy, we choose to be competitive with with state-of-the-art low-power NNs. Our target accuracy is to exceed $80\%$ , as typical solutions achieve $70-80\%$\cite{harvnet, multiexit_dac}. Processing time is kept under 1 second, suitable for the intended use case. By comparison, some ML accelerators achieve up to 20 s inference time at low power\cite{CNAv1}, while faster solutions (e.g., 0.2 s) exceed our power budget.

\begin{table}[t]
\captionsetup{
  justification = centering
}
\caption{Target Specification.}
\vspace{-.0in}
\centering
\scalebox{0.75}{\includegraphics{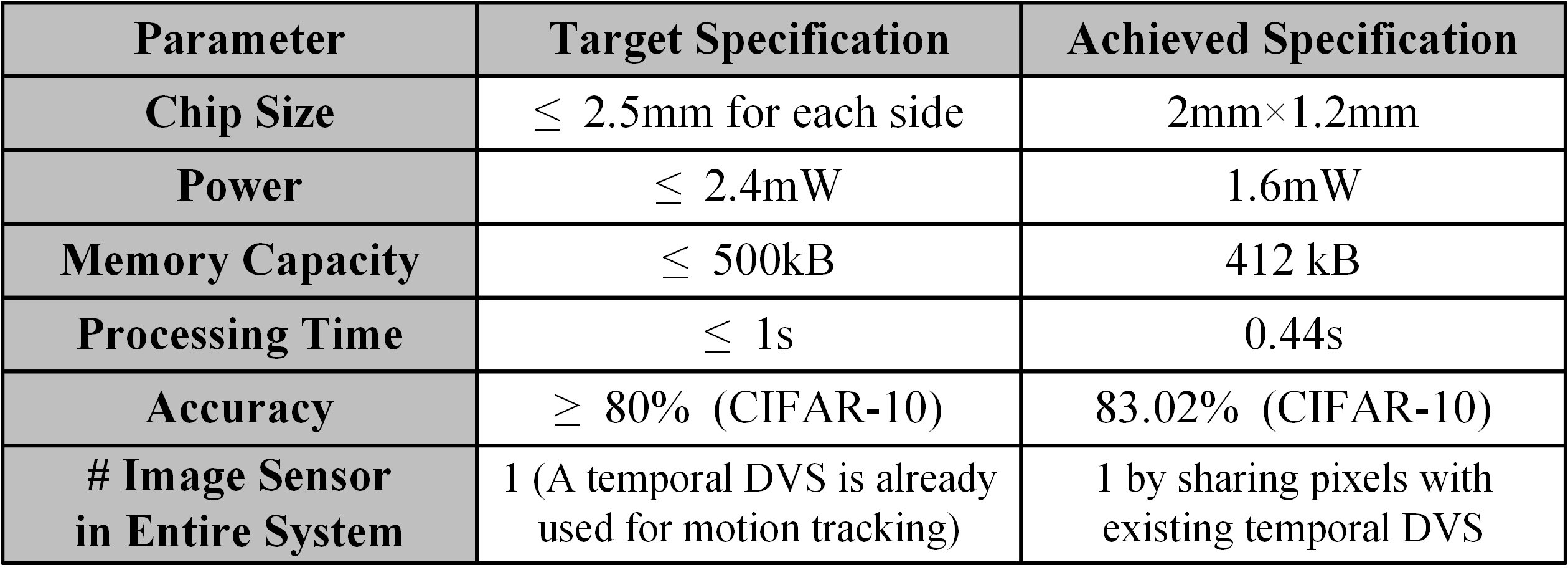}}
\vspace{-.3in}
\end{table}
\par\small To address the resource limitations in miniature systems, several approaches have been explored. First, sparsity-aware computing reduces computational complexity and power consumption by minimizing the number of MAC operations\cite{Seunghyun2024},\cite{Zhang21}. Secondly, binary neural networks (BNNs) and ternary-input binary-weight CNNs (TBNs) reduce memory usage by representing inputs, weights, and outputs using only 1 or 2 bits\cite{Seunghyun2024, Phil2021, Mauro2020, TBN}. TBN align well with DVS, as both operate with ternary values. A key advantage of DVS is the significant reduction in image data size\cite{DVS_temporal}, along with the elimination of power-hungry multi-bit analog-to-digital converter (ADC). While\cite{Fu2024} applies CNNs to DVS images, achieving reductions in both imager power and image size, it still fall short of meeting our full target specifications.

\par\small This paper proposes a TBN accelerator for millimeter-scale object classification systems using spatial DVS. The accelerator performs inference in 0.44 s, consumes 1.6 mW, and achieves a top-1 accuracy of $82.6\%$. Compared to prior CNN accelerators for miniature systems \cite{CNAv1}, our design improves the Figure-of-Merit (FoM) by a factor of 176. As shwon in Fig. 1(b), the key contributions are: (1) analysis of pixel combination strategies for generating DVS images to optimize recognition accuracy; (2) implementation the first hardware accelerator for TBN, to the best of our knowledge; and (3) integration of sparsity-aware zero-skipping, leveraging TBN's tenary nature to boost efficiency.

\par\small This work does not evaluate the efficiency of spatial DVS and TBN for high-complexity images, as they exceed the capabilities of the proposed system. Also, while CNNs on RGB images may achieve similar accuracy with greater efficiency, our approach prioritizes the use of DVS imagers to minimize overall system size by sharing pixels with the existing temporal DVS imager used for motion tracking.

\vspace{-.1IN} 
 \section{DVS Configuration \& TBN Justification}
\begin{figure}[!t]
\centering
\scalebox{0.75}{\includegraphics{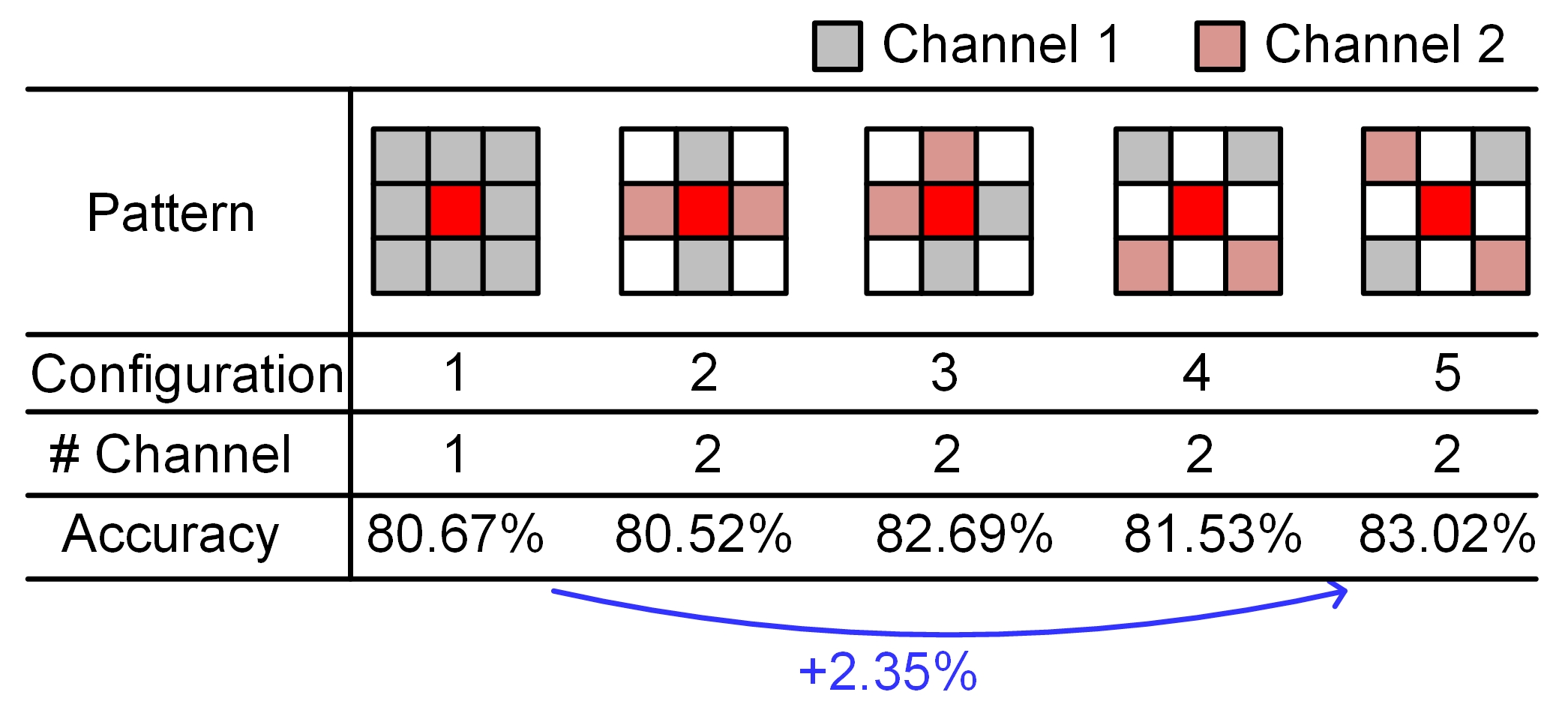}}
\vspace{-.1IN}
\DeclareGraphicsExtensions{.jpg}
\caption{Different DVS configurations and accuracy results.}
\vspace{-.2IN}
\label{fig_DVS_pattern}
\end{figure}
\begin{figure}[t]
\centering
\scalebox{0.75}{\includegraphics{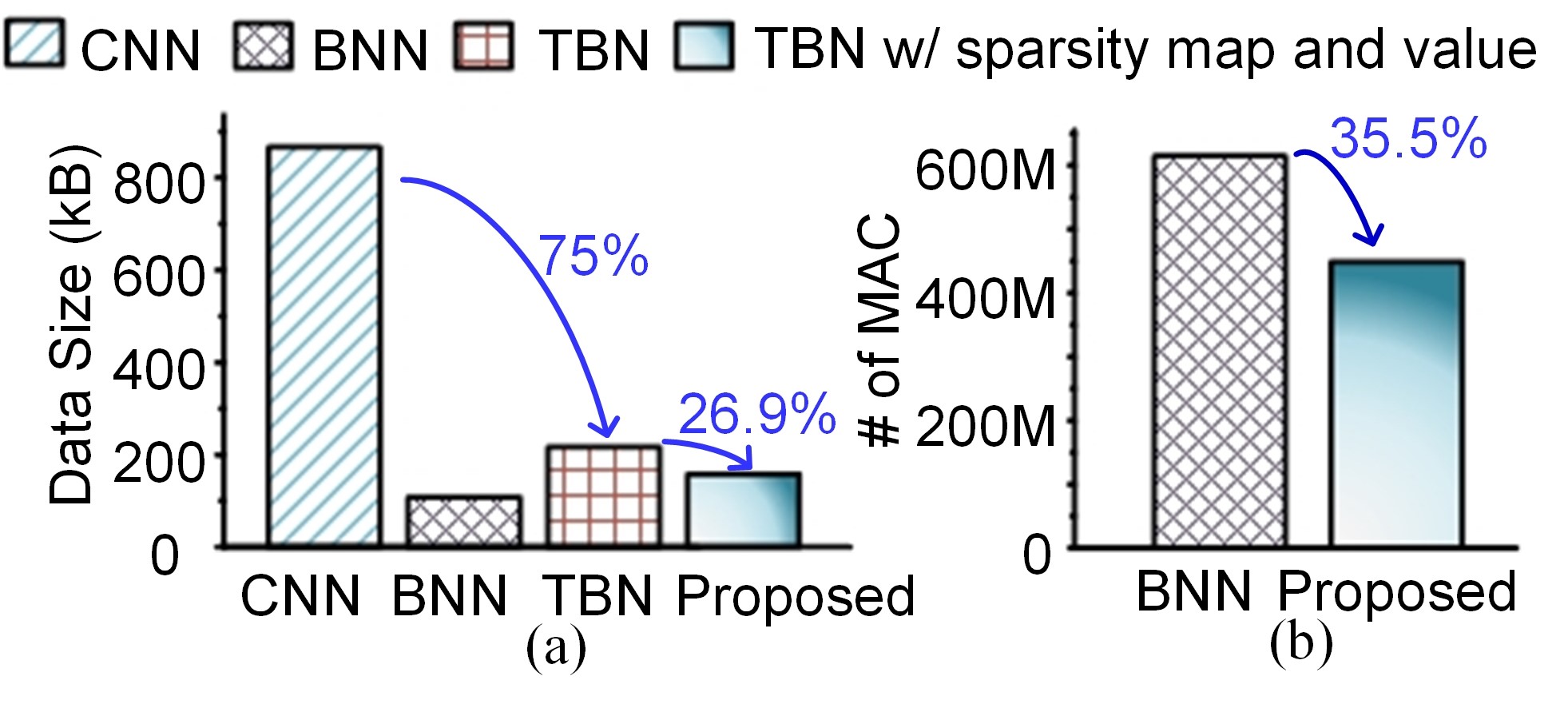}}
\vspace{-.1IN}
\DeclareGraphicsExtensions{.jpg}
\caption{DVS-based TBN's performance. (a) Data size reduction for the same NN. (b) MAC operation reduction via sparsity awareness.}
\vspace{-.3IN}
\label{fig_data_mac_comparison}
\end{figure}
\par\small Fig. 2(a) shows the spatial DVS working principle and its output. It compares a center pixel with its four diagonal neighbors, dividing them into two diagonal channels. Each channel averages the two diagonal pixels and subtracts this average from the center pixel. Two sets of comparators then compare the result to predefined thresholds. If the result is above the positive threshold, the output is +1; if below the negative threshold, the output is -1. These events update the corresponding cell in a 1-bit sparsity map to 1. The map resets to 0 at the start of each frame. In the value stream queue, +1 is stored as 1 and -1 as 0. Thus, zeros use only 1 bit in the sparsity map, while non-zero pixels require an additional bit in the value stream. Unlike simple thresholding, this DVS method compares neighboring pixels to detect edges, capturing more image information and compensating for uniform brightness.
\par\small Fig. 3 shows various spatial DVS configurations evaluated on the CIFAR-10 dataset using our TBN algorithm. Among the single-channel designs, Configuration $\#1$ achieves the highest accuracy. Configurations $\#2$ to $\#5$ include two output channels, with Configuration $\#5$ selected for the final design for its $2.35\%$ higher accuracy than Configuration $\#1$. Accuracy is a key factor as all configurations effectively reduced memory and power usage. Although more pixels per channel or extra output channels could improve accuracy, we do not pursue them due to their negative effects on fill factor and sensitivity. We assume the analog averaging and subtraction are implemented using capacitors, which further reduce fill factor and sensitivity.
\par\small In addition to spatial DVS, this work employs a TBN architecture. Fig. 2(b) illustrates a VGG-based TBN architecture consisting of six convolutional (CVx) layers, two fully connected (FCx) layers, three pooling and ReLU (PLx) layers, and three batch normalization (BNx) layers \cite{TBN}. Both convolution and fully connected layers use ternary inputs and binary weights. XOR gates replace multipliers, saving area and improving efficiency. Although both binary neural networks (BNN) and ternary-weight CNNs (TCN) can perform multiplication using XOR operations, this work utilizes TBN to reduce weight memory occupation, align with the ternary output of the target DVS as well as reduce the number of MAC operations further by skipping multiplication operations for zero inputs. At the end of certain convolutional, batch normalization, and fully connected layers, quantization (QNT) layers convert fixed-point data to ternary.
\par\small Fig. 4 quantifies the impact of using DVS input and TBN architecture. 
Compared to a conventional CNN, implementing the TBN architecture across all CVx and FCx layers reduces the feature map data size by $75\%$ by compressing 8-bit values into 2-bit representations, as is shown in Fig. 4(a). Additionally, the encoding method, which separates the sparsity map from the values, further decreases the size of the data by $26.9\%$. Compared to BNN, the sparsity-aware TBN reduces the number of MAC operations by $35.5\%$ by skipping multiplication or XOR operations for zero inputs, shown in Fig. 4(b). 
\vspace{-.1in}
 
 \section{Proposed TBN Accelerator Design}
\begin{figure}[!t]
\centering
\scalebox{0.80}{\includegraphics{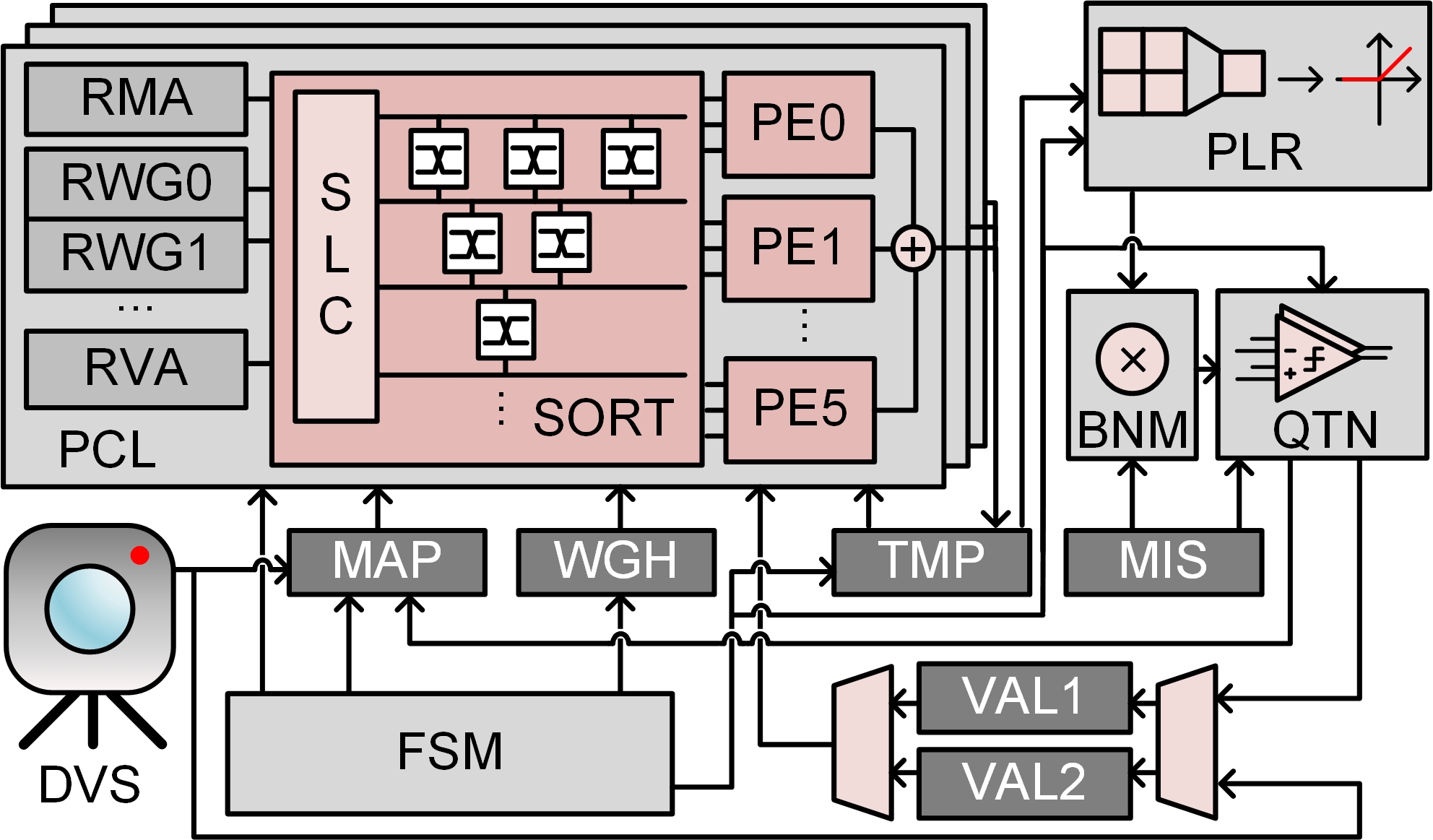}}
\DeclareGraphicsExtensions{.jpg}
\caption{Block diagram of the proposed accelerator.}
\vspace{-.2IN}
\label{fig: fig_block_diagram}
\end{figure}
\par \small Fig. 5 illustrates the block diagram of the accelerator designed for TBN. Separate SRAM blocks store the sparsity map (MAP), model weights (WGH), temporary partial sums (TMP), and miscellaneous values (MIS). Two value FIFOs (VAL1 and VAL2) hold the input and output value streams, respectively. Three processing clusters (PCLs) process the same row of the feature map while convolving with three rows of the weight kernel. Within each PCL, sparsity map registers (RMA), multichannel weight registers (RWG), and value registers (RVA) store data loaded from memory. A sorting network (SORT) balances workloads by reordering data grouped by the data slicer (SLC). Each PCL integrates six processing engines (PEs) to optimize area efficiency and processing time. A pooling-ReLU unit (PLR) processes the MAC results stored in TMP, and the batch normalization module BNM normalizes the pooling output by multiplying a pretrained factor. The PLR and BNM process multibit fixed-point data while a quantization unit (QTN) uses multiple comparators to compare output against predefined thresholds stored in MIS, encoding the results into the sparsity map and value stream. The accelerator processes the TBN layer by layer, storing the input and output feature maps, as well as the weights of each layer, in the on-chip memory. Each layer's image size, channel count, and options for pooling and batch normalization are configurable. An FSM-based controller (FSM) receives these parameters and manages the memory data flow accordingly. 
\begin{figure}[!t]
\centering
\scalebox{0.80}{\includegraphics{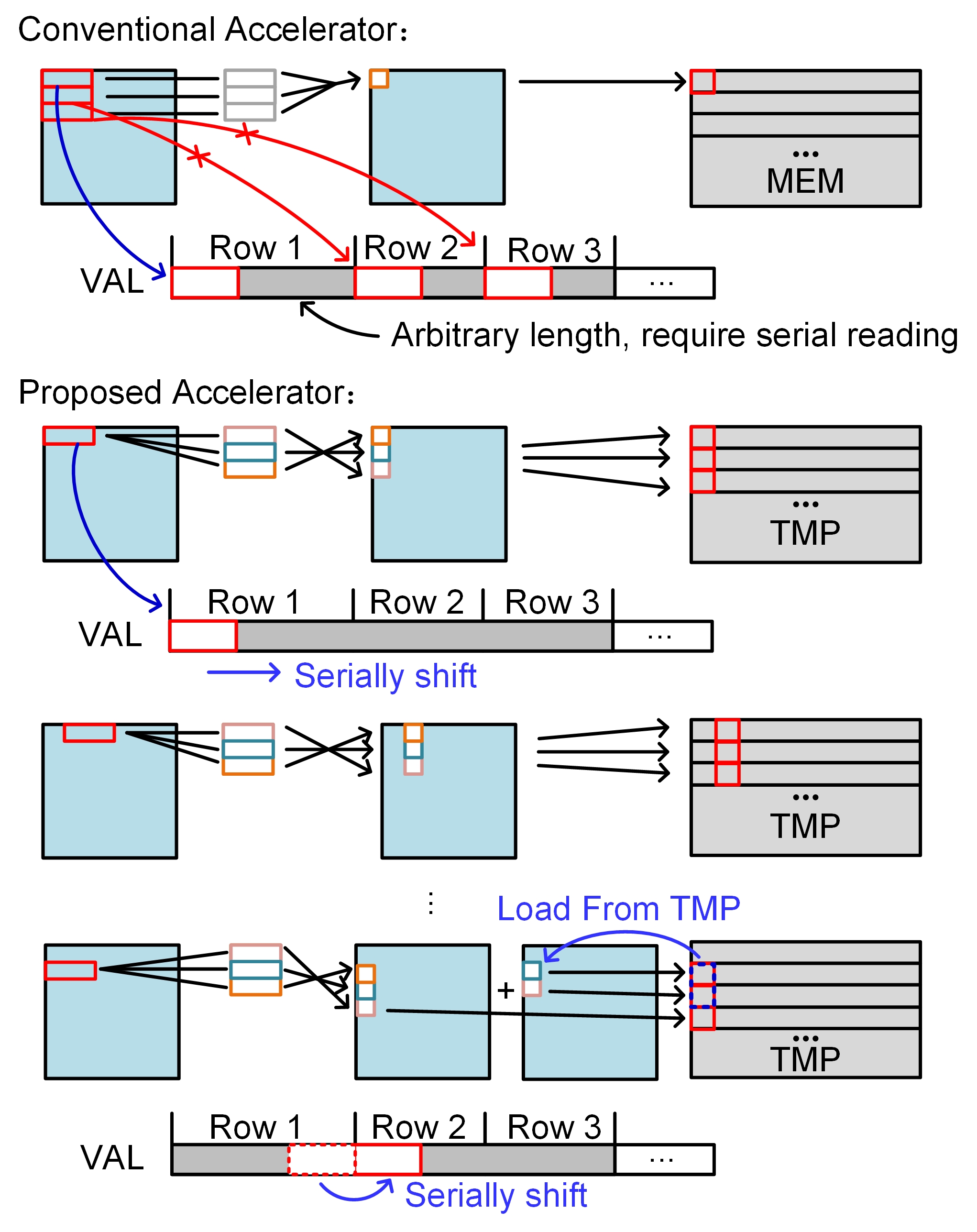}}
\DeclareGraphicsExtensions{.jpg}
\vspace{-.1in}
\caption{Processing order in conventional and proposed accelerators.}
\vspace{-.3in}
\label{fig_processing_order}
\end{figure}
\par\small Fig. 6 compares 3×3 convolution processing in the PCL of the proposed accelerator with a conventional method. A conventional accelerator maps multiple rows of the feature map directly to the 2D kernel window \cite{CNAv1}. As shown in Fig. 6 (top), it loads a 3×3 feature map block, multiplies it with the kernel, and sums the results. In contrast, the proposed TBN uses sparsity-aware processing, which handles values stored at irregular memory positions. These values cannot be accessed simultaneously, as each row contains data at arbitrary locations. Therefore, the VAL is designed as a FIFO, reading values sequentially based on the sparsity map. This means the second row cannot be processed until the first row's values are fully read. To solve this, the proposed accelerator processes a 1×3 input window at a time (Fig. 6, bottom). It generates three partial sums by convolving the input with each row of the kernel. These partial results are stored in TMP. The input window then shifts right to read the next values. When the window reaches the boundary and moves to the next row, the stored partial sums are reloaded into the PCL and added to the new sums to produce the final output.
\par\small Fig. 7 illustrates how the data are stored and loaded. As shown in Fig. 7(a), MAP holds the sparsity map data in a channel-first order \cite{Zhang21}. In 32-bit MAP memory, elements from channels 0 to 31 in the top-left corner of the feature map are stored at the same address, followed by 32 channels in the top-second-left position and so on. After the first 32 channels of all 1024 elements are aligned in memory, the next address holds data from the same position on the next set of 32 channels (channels 32–63). Weights are stored in WGH in a similar format, with each row flattened. The VAL FIFO holds only the non-zero values. Fig. 7(b) shows during convolution or fully connected operations, feature map values are broadcast to the RMA and RVA registers across all PCLs. At the same time, weight data for each row is sent to one corresponding PCL.
\begin{figure}[!t]
\centering
\scalebox{0.80}{\includegraphics{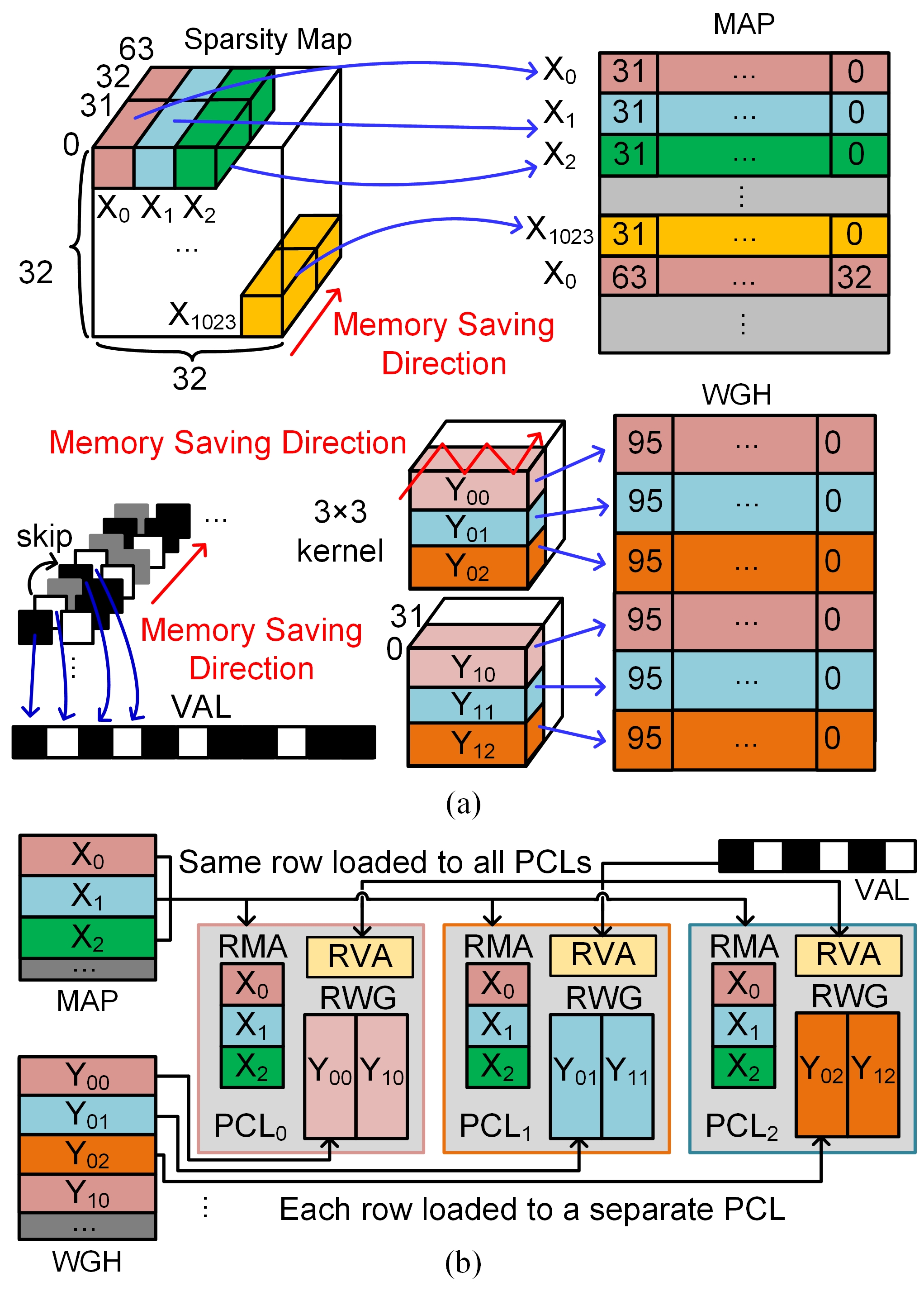}}
\vspace{-.1in}
\DeclareGraphicsExtensions{.jpg}
\caption{Data format of the sparsity map, values, and weights in memory. (a) Data storage format. (b) Data loading.}
\vspace{-.2in}
\label{fig_memory_format}
\end{figure}
\begin{figure}[t]
\centering
\scalebox{0.8}{\includegraphics{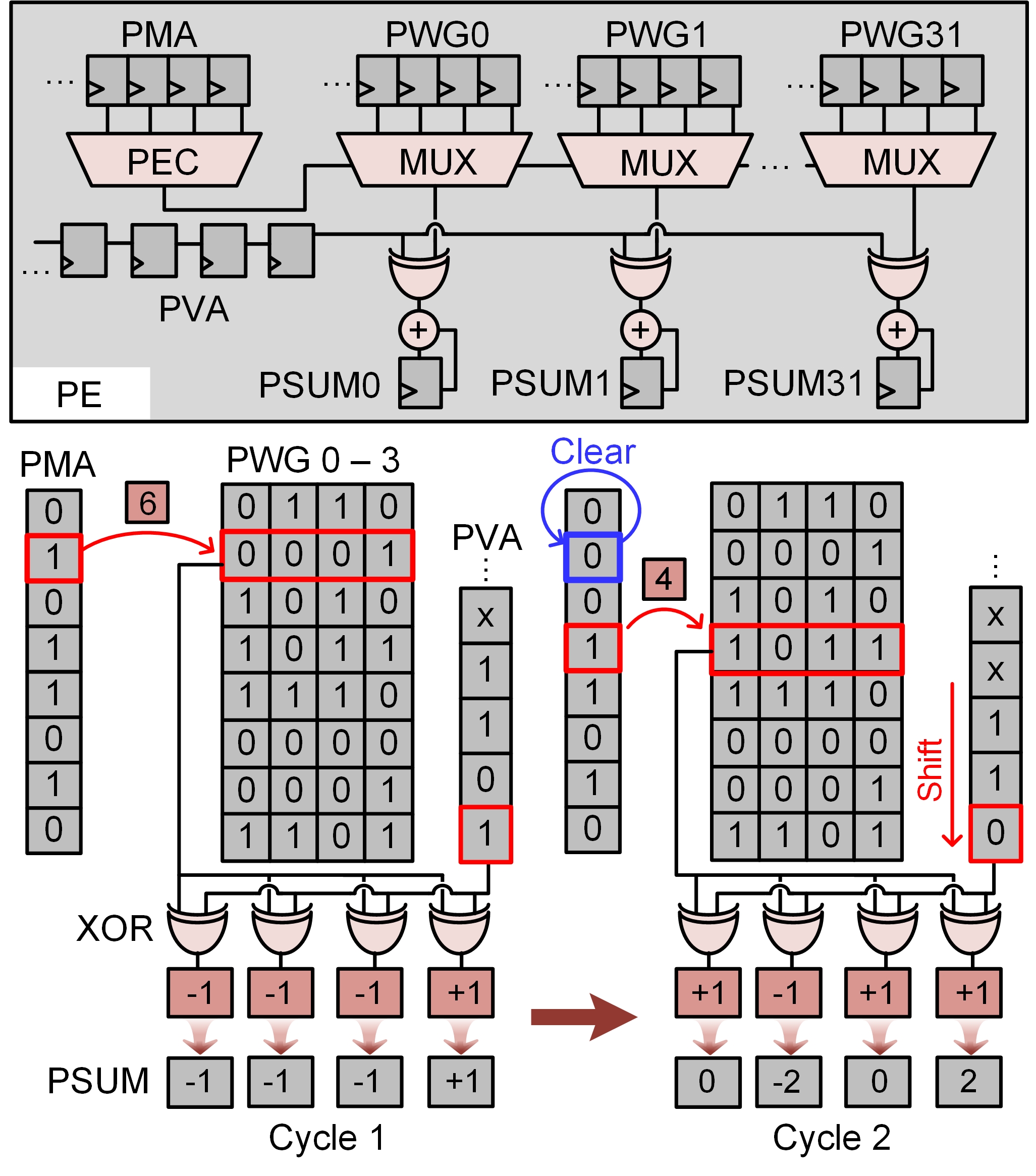}}
\vspace{-.1IN}
\DeclareGraphicsExtensions{.jpg}
\caption{PE design and convolution process.}
\vspace{-.2IN}
\label{fig_processing}
\end{figure}

\par\small Fig. 8 illustrates the MAC operation in the PE of the PCLs. In Fig. 8 top, registers PMA, PVA, and PWG hold parts of RMA, RVA, and RWG, respectively. A priority encoder (PEC) identifies the highest active bit in PMA and sends this position to a multiplexer (MUX), which selects the corresponding data for multiplication using XOR. The PVA register shifts one bit at a time, and 32 XOR gates perform the multiplication. The results are accumulated in the PSUM register. Fig. 8 bottom shows an example. The PEC first detects the 6th bit in PMA. The MUX selects the 6th row from PWG and XORs it with the first bit in PVA. The result updates PSUM. Then, PVA shifts right by one bit. After clearing the 6th element in the PMA, the PEC skips the 5th bit and detects the next ‘1’ at the 4th bit and the process continues.
\begin{figure}[t]
\centering
\scalebox{0.8}{\includegraphics{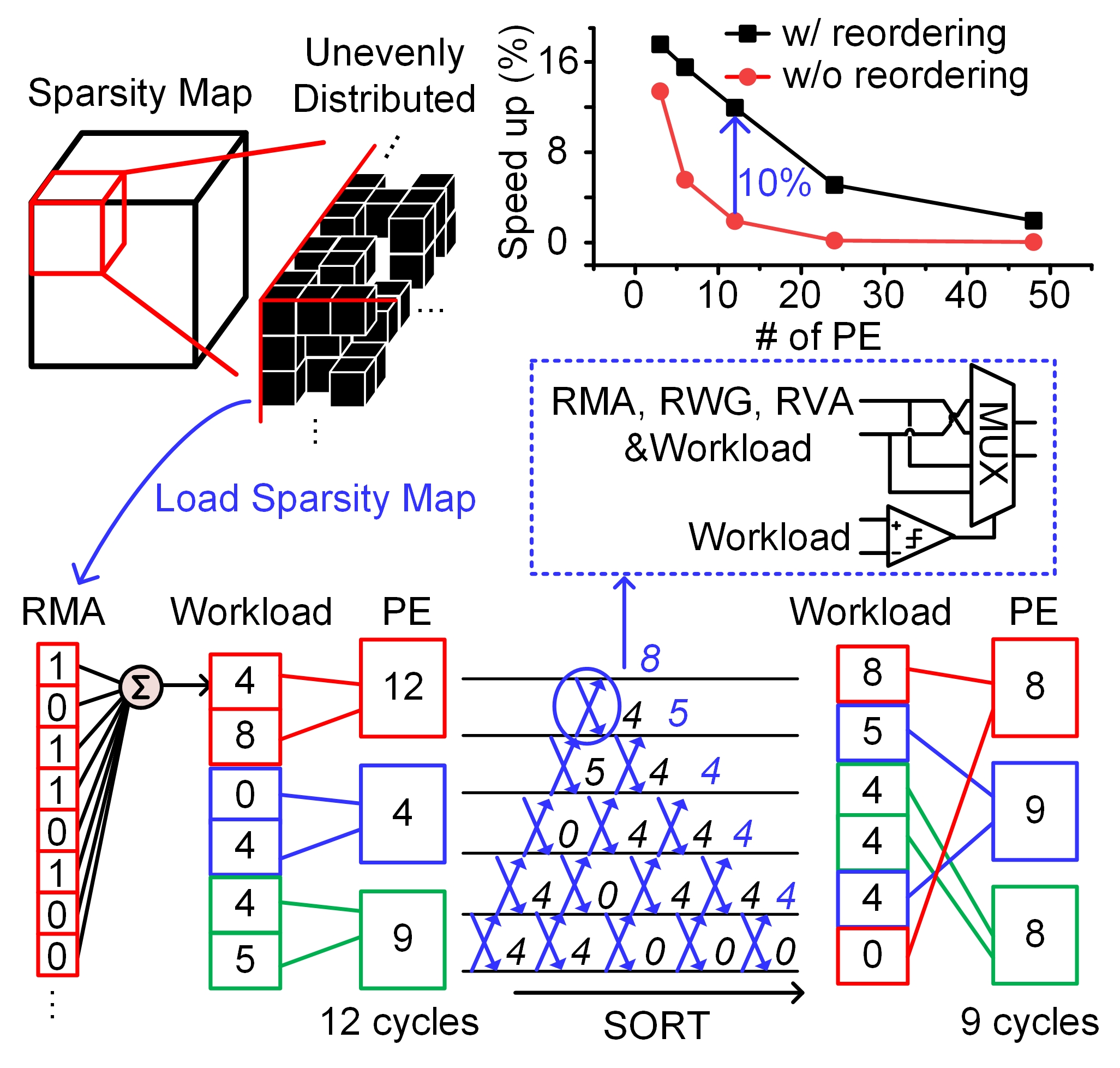}}
\vspace{-.1in}
\DeclareGraphicsExtensions{.jpg}
\caption{Unevenly distributed elements in the sparsity map (top left), workload balancing using a sorting network (bottom), and speed up by sparsity-aware zero-skipping w/ and w/o reordering (top right).}
\vspace{-.2IN}
\label{fig_reorder}
\end{figure}
\par\small Due to multiple PEs processing parts of the workload simultaneously, the random distribution can lead to uneven processing time. Fig. 9, using a 3-PE and 16-channel per PE model for simplified illustration, shows that with the original workload distribution, the first PE receives 12 non-zero values (workload) which require 12-cycles for the MAC operation. The second PE only receives 4 values and remains idle for the rest 8 cycles. To reduce idle time and improve general PE utilization, SORT redistributes workload more evenly in a single cycle. Input data are split into groups twice as much as the number of PEs (E.g., 6 groups for 3 PEs and 12 groups for 6 PEs). Then, the sorting network reorders the groups according to the amount of non-zero values. Once the data are sorted, the PEs take one highest and one lowest workload from the unassigned tasks. In the presented example, the worst-case processing time is reduced to 9 cycles, reducing the processing time by $25\%$. Compared to the baseline test case (without 0-skipping and reordering), simply adding more PEs does not effectively reduce processing time due to the bottlenecked PE. However, optimizing workload balance reduces processing time by up to $10\%$ with 6 PEs. Under 10 MHz clock frequency, this configuration achieves 46.4 GOPS processing throughput for CIFAR-10 dataset and the TBN architecture in Fig. 2 (b), higher than the state-of-the-art design\cite{CNAv1} while maintaining a low area cost.
\par\small Fig. 10 illustrates how the QNT block handles ternary inputs and outputs across layers. The QNT block contains logic similar to the comparison block in the DVS generation process. After the PCL completes the MAC operation, the partial sums are read from TMP. The integrated logic then compares these partial sums with a pre-trained threshold value and encodes the results back into the sparsity map and values.
\begin{figure}[t]
\centering
\scalebox{0.80}{\includegraphics{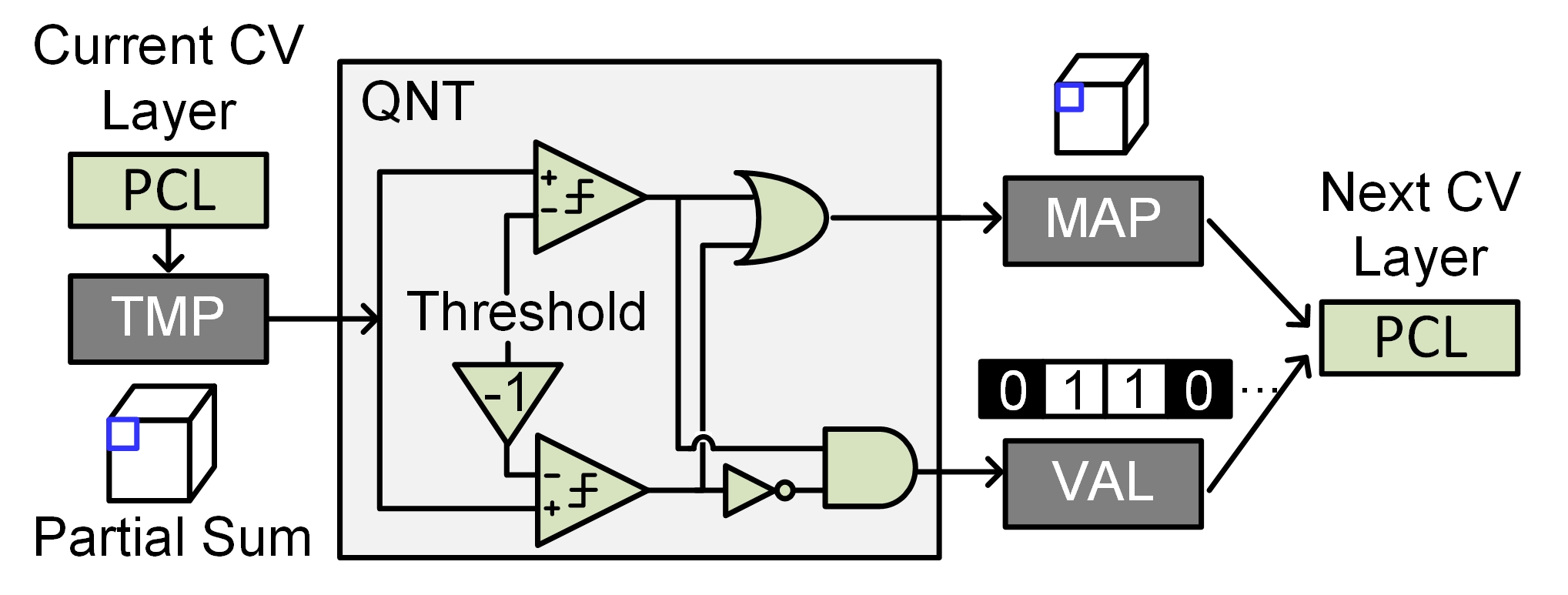}}
\DeclareGraphicsExtensions{.jpg}
\vspace{-.1in}
\caption{Quantization process after each convolution layer.}
\vspace{-.2IN}
\label{fig_QNT}
\end{figure}
\vspace{-.1IN}

 \section{\small Experimental Results}
\begin{figure}[!t]
\centering
\scalebox{0.75}{\includegraphics{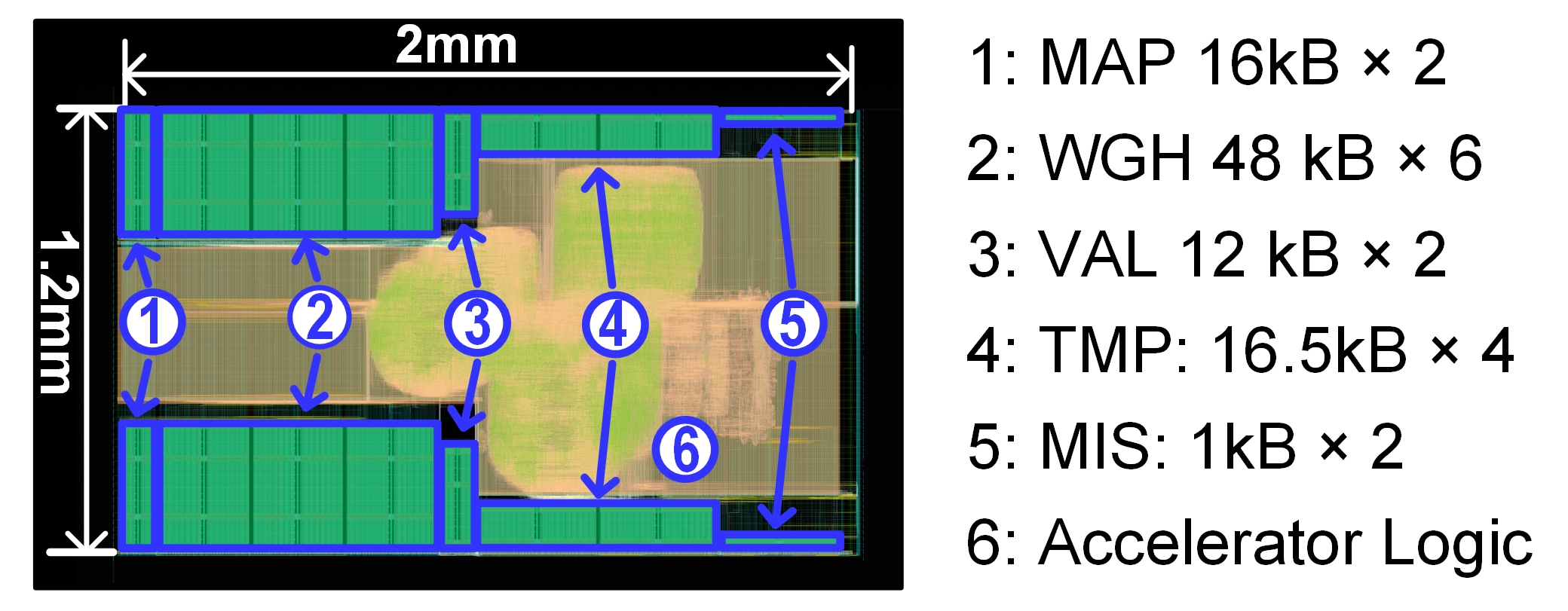}}
\DeclareGraphicsExtensions{.jpg}
\vspace{-.1in}
\caption{Layout of the implemented TBN accelerator.}
\vspace{-.2IN}
\label{fig_layout}
\end{figure}
\begin{figure}[!t]
\centering
\scalebox{0.75}{\includegraphics{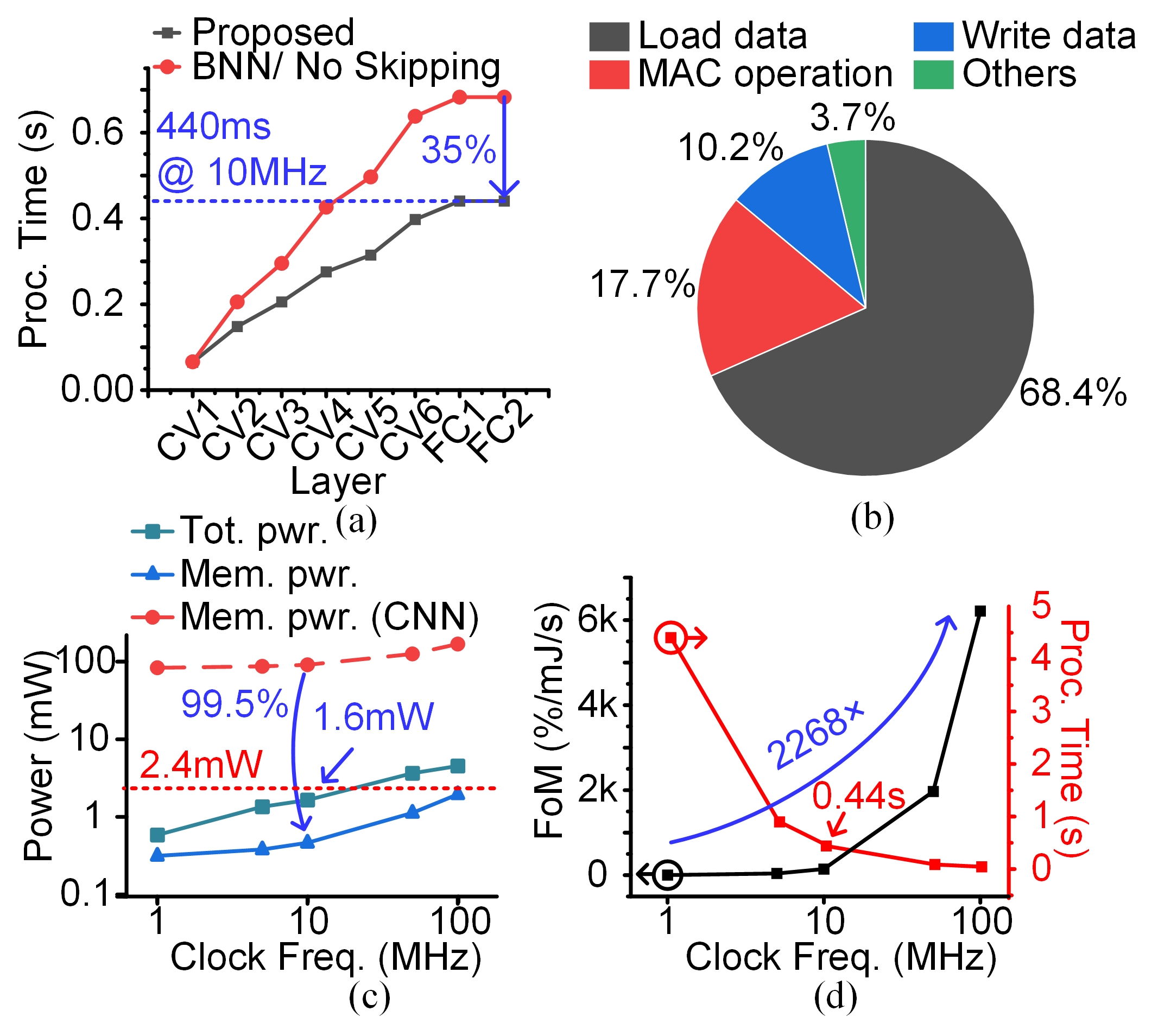}}
\DeclareGraphicsExtensions{.jpg}
\vspace{-.1in}
\caption{Simulated performance of the proposed design. (a) Accumulated processing time across TBN layers. (b) Breakdown of processing time. (c) Power consumption across frequencies. (d) Processing time and FoM across frequencies.}
\vspace{-.3in}
\label{fig_measurements}
\end{figure}
\begin{table*}[ht]
\captionsetup{
  justification = centering
}
\caption{Performance Comparison For NN Accelerators Featuring Power Efficient, Sparsity Aware or BNN/TBN.}
\centering
\scalebox{0.75}{\includegraphics{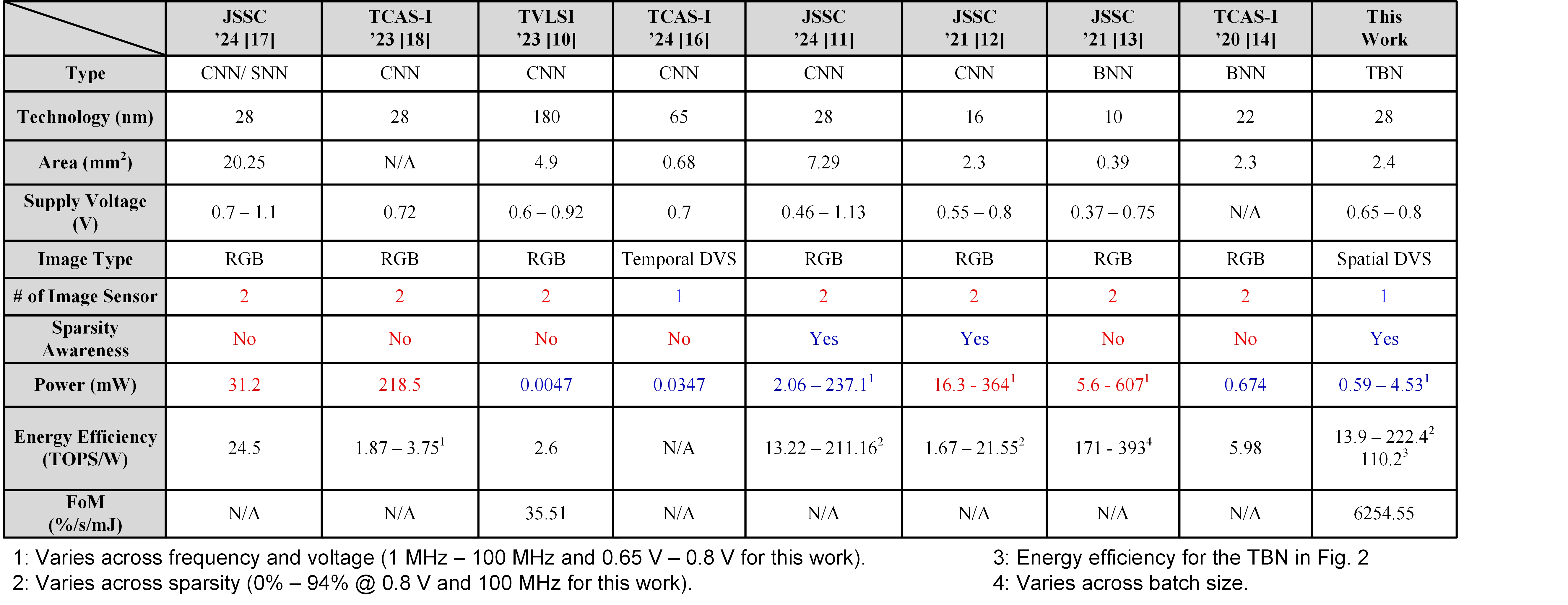}}
\vspace{-.3in}
\end{table*}
\par\small The proposed design is implemented in a 28 nm CMOS process. Fig. 11 shows the layout, which occupies an area of 2.4 mm². Since each part of the TBN requires different data widths, multiple memory blocks with varying widths are integrated. The total on-chip memory is 412 kB. The top-1 inference accuracy achieved is $82.56\%$, using the CIFAR-10 dataset converted to the selected DVS output format.
\par\small Fig. 12 shows the simulation results. Fig. 12(a) presents the processing time defined as the total cycles per layer divided by the frequency at a 10 MHz clock frequency. The average processing time is reported due to random sparsity. Compared to an equivalent BNN, the proposed design reduces the processing time by $35\%$ during inference. Fig. 12(b) breaks down the processing time, revealing that data fetching from memory dominates the total time since the values being read out serially from VAL FIFO. The typical dominant MAC operation only accounts for $17.7\%$ of the processing time in the proposed accelerator, thanks to zero skipping and workload reordering.
\par\small Fig. 12(c) depicts the power consumption across frequencies ranging from 1 MHz to 100 MHz. The power consumption of the accelerator is simulated at the TT corner using the Synopsys CustomSim Simulator with a precision level of 5. Due to the significant data size reduction, memory power is reduced by $98.8\%$ compared to a CNN accelerator running the same network architecture. 
At 10 MHz, the proposed design consumes 1.6 mW, which represents $67\%$ of the total system power budget.
\par\small Fig. 12(d) shows the processing time and Figure-of-Merit (FoM) across frequency, defined as:
\begin{equation}
    FoM = \frac{Accuracy}{Processing\;Time \times Energy}
\end{equation}
The operating clock frequency increasing from 1 MHz to 100 MHz leads to a significant reduction in processing time and energy consumption, resulting in a $1291\times$ improvement in FoM. Under 10MHz clock frequency, the system completes one inference in 0.44 seconds. The benefit of more parallel MAC units and faster clock speed overdrives the increased power, leading to an $7.26 \times$ of the FoM ($257.9 \%/s/mJ$) improvement over previous CNN inference accelerators for miniature systems \cite{CNAv1}.
\par\small Table II compares the performance of the proposed design with state-of-the-art power-efficient, sparsity-aware, and BNN accelerators. While the cited designs are based on silicon measurements, the proposed design is evaluated using SPICE simulation. To assess suitability for miniature sensing tasks, we examine the specifications in Table I for each work. Among the cited designs,\cite{Kim2024, Xie2023,  Zhang21}, and\cite{Phil2021} consume more than 2.4 mW of power, exceeding the limits of the miniature battery used in our target system. Works\cite{CNAv1} and\cite{ Seunghyun2024} accelerate CNN computations for object recognition using RGB imagers. However, supporting these systems would require an additional image sensor alongside the temporal DVS imager in our target system, increasing area and complexity, both critical constraints. Furthermore, CNN-based designs require multi-bit multipliers, which reduce power efficiency when staying within an acceptable power budget. Work\cite{Mauro2020} improves efficiency by employing BNNs, but lacks support for sparsity, limiting overall energy efficiency. Meanwhile, \cite{Fu2024} processes temporal DVS images, allowing the same sensor to be used for motion tracking and eliminating the need for a separate image sensor for object detection. However, their accelerator is optimized for compact CNNs trained on hand gesture recognition datasets, which are significantly simpler than the CIFAR-10 models used in our work. For a fair comparison, \cite{Fu2024} would need to be re-evaluated using a dataset similar in complexity to CIFAR-10.

\section{\small Conclusion}
\vspace{-0.5PT}
\par\small This work introduces a hardware accelerator for millimeter-scale object classification systems, combining a spatial DVS with a TBN architecture. The proposed design enables inference within 440 ms, operates at a power consumption of 1.6 mW, and achieves a top-1 accuracy of $82.6\%$. Compared to existing CNN accelerators for miniature systems, this approach enhances the FoM by a factor of $7.3 \times$.
\vspace{-.1in}

\bibliographystyle{IEEEtran}
\bibliography{tex/Reference}

\end{document}